\documentstyle[epsf,12pt]{article}
\newcommand{\beq}{\begin{equation}}
\newcommand{\eeq}{\end{equation}}
\newcommand{\bea}{\begin{eqnarray}}
\newcommand{\eea}{\end{eqnarray}}
\newcommand{\barr}{\begin{array}}
\newcommand{\earr}{\end{array}}
\newcommand{\bmat}{\left( \begin{array}}
\newcommand{\emat}{\end{array} \right)}

\newcommand{\bit}{\begin{itemize}}
\newcommand{\eit}{\end{itemize}}

\newcommand{\rttwo}{\sqrt{2}}
\newcommand{\id}{{\sf{'}}{\hspace{-0.8mm}|\hspace{-0.6mm}|}\hspace{-1.6mm}
\rule[-1mm] {1.5mm}{.1mm}\hspace{-1.1mm}\rule[3.0mm]{.65mm}{.1mm}}
\newcommand{\B}{{^8}\!B}
\newcommand{\Be}{{^7}\!Be}
\newcommand{\res}{(N_e)^{res}}
\newcommand{\ncr}{(N_e)^{cr}}
\newcommand{\nmax}{(N_e)_{max}}
\newcommand{\Psurv}{\bra P(\nu_e \ra \nu_e) \ket}
\newcommand{\bra}{\langle}
\newcommand{\ket}{\rangle}

\newcommand{\nue}{\nu_e}
\newcommand{\nm}{\nu_{\mu}}
\newcommand{\nt}{\nu_{\tau}}
\newcommand{\temu}{{\theta}_{12}}
\newcommand{\tetau}{\theta_{13}}
\newcommand{\tmt}{\theta_{23}}
\newcommand{\th}{\theta}
\newcommand{\thm}{\theta^m}
\newcommand{\ra}{\rightarrow}
\newcommand{\impl}{\Longrightarrow}
\newcommand{\df}{\Delta f}

\newcommand{\dm}{\Delta m^2}

\begin{document}
\begin{titlepage}

\title{Mass or Gravitationally Induced Neutrino Oscillations? --
A Comparison of $\B$ Neutrino Flux Spectra
in a Three--Generation Framework}

\vspace{10pt}
\author{J.\ R.\ Mureika\thanks{newt@avatar.uwaterloo.ca} ~
and
R.\ B.\ Mann\thanks{mann@avatar.uwaterloo.ca}\\
\\
{\it Department of Physics} \\
{\it University of Waterloo} \\
{\it Waterloo, Ontario N2L 3G1  Canada}}
\maketitle

\begin{abstract}
Both gravitational and mass induced neutrino oscillation mechanisms provide
possible resolutions to the Solar Neutrino Problem.
The distinguishing feature between the
two mechanisms is their dependence on the neutrino energy.
We investigate the implications of this by computing
 the $\B$ neutrino spectrum as determined from each mechanism
using a realistic three--flavor evolution model.  We
find that in the limit of small $\tetau$ mixing angle, the
differences are significant enough to observe in future
solar neutrino experiments.
\end{abstract}

\end{titlepage}
\setcounter{page}{1}

	The Solar Neutrino Problem (SNP) has perplexed
both astrophysicists and particle physicists for upwards of thirty
years now.  The measured flux of $\nue$s  incident on the Earth--bound
detectors \cite{clx,kii,galx,sage} remains in conflict with
the standard solar model prediction (see \cite{bah2} for a review).  A
plausible
expanation of this discrepancy is that oscillations take place between
electron neutrinos and neutrinos of differing species, thereby reducing
the expected $\nue$ flux to an empirically acceptable value.

At present there are two qualitatively distinct mechanisms which could
give rise to such oscillations.  One is the well-known
Mikayev--Smirnov--Wolfenstein (MSW) mechanism \cite{msw1,msw2,wolf}
which postulates that
neutrinos possess a non--trivial mass eigenbasis, in contrast to the
assumptions of the Minimal Standard Model. In this mechanism electron
neutrinos produced at the core of the sun will,
under certain conditions dependent upon the solar electron density,
undergo a resonance with other species of neutrinos whose flux then
has no effect on earth-based detectors.  Another mechanism proposed
more recently \cite{gasp} hypothesizes that neutrinos posess
a flavor--dependent coupling to the external
gravitational field.  This mechanism (recently dubbed the VEP mechanism
\cite{bah1}) violates the Einstein Equivalence Principle (EEP), since
it requires $G_i = (1+f_i) G$, where $G$ is Newton's constant, $i$ a
flavor index, and $f_i$ dimensionless parameters which characterize
the degree of EEP violation,
with each $f_i \ll 1$.  To ensure the full effect of three flavors,
we must have $f_i \ne f_j, i \ne j$.  For first generation neutrinos,
we define $f_1 = 0$, {\it i.e.} $G_1 = G$.
The VEP mechanism does not require neutrinos to have
a non-degenerate mass-matrix.

Both oscillation mechanisms rest on the assumption that
the two neutrino eigenstates, flavor $\left(|\nu\ket_W\right)$ and
mass/gravitational $\left(|\nu\ket_{M,G}\right)$, are
related by an $SU(N_g)$ transformation,
\beq
|\nu \ket_W = V_3 \, |\nu\ket_{M,G}~,
\label{rot}
\eeq
where $N_g=3$ for three flavors.  These states evolve according
to the equation \cite{gasp}
\bea
 & i\frac{d}{dr}\,|\nu\ket_{M,G} = H_{M,G} \, |\nu\ket_{M,G}\nonumber
\\ ~\impl~ &
i\frac{d}{dr}\, |\nu \ket_W =\left\{ V_3^{\dagger}
H_{M,G} V_3+A(r)\right\}  \, |\nu \ket_W~.
\label{evol}
\eea
Flavor oscillations and resonances arise due to the off--diagonality
of the modified Hamiltonian $V_3^{\dagger} H_{M,G} V_3+A(r)$, where
$A(r) = diag(\rttwo G_F N_E(r),0,0)$ is the term in the Hamiltonian
corresponding to  $\nue$-$e$
({\it i.e.} charged--current) electroweak interactions.
If we assume no  CP--violation in the neutrino
sector, then the four--parameter matrix $V_3$ reduces to a real,
orthogonal rotation with three vacuum mixing angles $\temu,\tetau$, and $\tmt$.

	The main difference between the MSW and VEP mechanisms manifests
itself in the energy dependence of the evolution equations in (\ref{evol}).
which respectively are
\bea
H_{M}& =& \frac{1}{2E} \bmat{ccc} m_1^2 & 0 & 0 \\ 0 & m_2^2 & 0 \\
0 & 0 & m_3^2 \emat~, \\
\label{mswham}
H_{G} & = & 2E|\phi| \bmat{ccc} f_1 & 0 & 0 \\ 0 & f_2 & 0 \\
0 & 0 & f_3 \emat~,
\label{vepham}
\eea
where a factor of unity has been subtracted out in each of (\ref{mswham})
and (\ref{vepham}) as it contributes only an overall unobservable phase.
We have taken this factor to be $\id \cdot (H_{M,G})_{11}$, leaving the
dynamics of the mechanism dependent of the eigenvalue differences
$\dm_{ij} \equiv m_i^2 - m_j^2~,~\df_{ij} \equiv f_i - f_j$.

	In this paper we consider the effect of these
differing energy dependences on the suppression of the $\B$ neutrino
flux in the three--generation scheme.  We show that for
a small $\nue \ra \nt$ mixing angle $\tetau$, there is a noticeable
energy dependence in the shape and size of the suppression curves
for MSW and VEP.  This  could be easily detected in present--day
and future water detectors.  For large  $\tetau$, the suppression
is energy--independent, and hence there is little variation
in the spectrum. We shall take $\phi \equiv \phi_{\odot}(r)$ in
(\ref{vepham}), where $\phi_{\odot}(r)$ is the solar gravitational
potential\footnote{We do not consider the effects of
$\nue$--regeneration in the Earth (the ``day--night effect''),
nor do we consider the minimal contribution of the local supercluster.
For discussions as to why these can be excluded, see \cite{bah1}, or
\cite{jrm1}.}.

In order to study the  suppression of $\nue$s, we must
determine the model--dependent survival probability for these
neutrinos as they travel from the solar center to the Earth--based
detectors.  Averaged over 1 AU, the solution to (\ref{evol})
can be written \cite{bald}
\bea
\Psurv& =& \sum_{i,j=1}^{3} |(V_3)_{1i}|^2\; |(P_{LZ})_{ij}|^2\;
|(V_3^m)_{1j}|^
2 \nonumber \\
 & = & c_{m12}^2 c_{m13}^2 \left\{ (1-P_1)c_{12}^2
 c_{13}^2 + P_1 s_{12}^2 c_{13}^2 \right\} \nonumber \\
& & + s_{m12}^2 c_{m13}^2 \left\{ P_1 (1-P_2) c_{12}^2 c_{13}^2 +
(1-P_1)(1-P_2)
s_{12}^2 c_{13}^2 + P_2 s_{13}^2 \right\} \nonumber \\
& & + s_{m13}^2 \left\{ P_1 P_2 c_{12}^2 c_{13}^2 + P_2 (1-P_1) s_{12}^2
s_{13}^2 +(1-P_2) s_{13}^2 \right\}~. \nonumber \\
{  }
\label{p3}
\eea
The terms
\beq
 s_{ij} \equiv \sin{\th_{ij}}~,~c_{ij} \equiv  \cos\th_{ij}
\eeq
refer to the vacuum mixing angles, while $s_{mij},c_{mij}$ are the analogous
matter--enhanced  trigonometric functions \cite{zag}.  The referenced
work expresses the functions in terms of MSW parameters; for VEP,
we make the global substitution
\beq
\frac{dm_{ij}}{2E} ~\ra ~ 2E|\phi| \df_{ij}~.
\label{subs}
\eeq
The functions $P_i$ are the Landau--Zener
jump probabilities \cite{bald} for non--adiabatic state transitions, and are
inherently energy--dependent.

The matter--enhanced
mixing angles $\thm_{1j}$ \cite{zag} are dependent upon the point of production
of
the solar $\nue$.  If this is above the resonance density $\res_{1i}$ for the
specific 12-- or 13--flavor transition ({\it i.e.} $\nue \ra \nm$ or
$\nue \ra \nt$), then $\thm_{1i} \ra \frac{\pi}{2}$.  Hence, the
probability in (\ref{p3}) reduces to
\beq
\Psurv = P_1 P_2 c_{12}^2 c_{13}^2 + P_2 (1-P_1) s_{12}^2 s_{13}^2
+(1-P_2) s_{13}^2~,
\label{lim1}
\eeq
For large $\tetau$, the jump probabilities $P_2$ vanish ({\it i.e.} adiabatic
approximation for 13--transitions), and we are left with
\beq
\Psurv =  s_{13}^2~,
\label{lrg}
\eeq
which is energy--independent (and, interestingly enough, independent
of the 12--transitions).  Conversely, the small $\tetau$ limit of
(\ref{lim1}) is
\beq
\Psurv = c_{12}^2 P_1 P_2~
\label{sm}
\eeq
which shows distinct energy dependence on both 12-- and 13--flavor
transitions.  By studying the two limits in question, it is possible
to see exactly how the third flavor affects the dynamics of the
oscillation mechanism.  The small $\tetau$ solutions  will in
some respects approximate the two--flavor mechanism (which is
fully recovered in the limit $\tetau \ra 0$).  At the other extreme, it
is possible that the large $\tetau$ solutions can yield spectral
distortions which match observed $\B$ fluxes, but cannot be
obtained via the two--flavor model.

	Since it was stated that the main difference
between the VEP and MSW oscillation mechanisms should be visible in a
study of the energy--dependent suppression of the neutrino flux, we
show that striking differences can be observed for certain input parameters
of the two models.

Although the $\B$ decay is one of the rarest nuclear reactions in the Sun,
the resulting $\B$ neutrinos are the easiest to study,  since
they have the widest spectrum of $E \in [0,15]\:$MeV, and
are of sufficiently high energy to be observed by the
$\nue$-$e$ scattering detectors ({\it e.g.} Kamiokande II, Superkamiokande,
SNO).  Hereafter, we provide a comparison of the $\B$ flux
curves as reduced by VEP and MSW, for both natural ($\df_{31} > \df_{21},
 \dm_{31}>\dm_{21}$) and broken eigenvalue--hierarchies ($\df_{31}<\df_{21},
 \dm_{31}<\dm_{21}$).

	While chemical detectors can ascertain the rate of solar neutrinos
which reach the Earth, the water--based detectors provide extra pieces
of information which can help pin down parameters.  An accurate measurement
of the incident $\B$ flux can tell us such things as which direction
the neutrinos came from, exact arrival times (assuming they originate from
the Sun) and, most importantly, show the energy--dependence of the
suppression mechanism at work.  Due to the fact that the \u{C}erenkov
detectors have relatively high energy thresholds, neutrinos whose energies
are below these thresholds
will not be visible.  For example, KII has a threshold energy
of $E_{th} = 9\:$MeV \cite{bah3}, while SNO will have a  lower one of
$E_{th} \approx 5\:$MeV.  Thus, these experiments will be particularly
useful if the parameter sets are such that
suppression is visible in the high energy portion of the $\B$ spectrum.

\begin{figure}
\leavevmode
\epsfysize=500pt
\epsfbox[60 20 70 470] {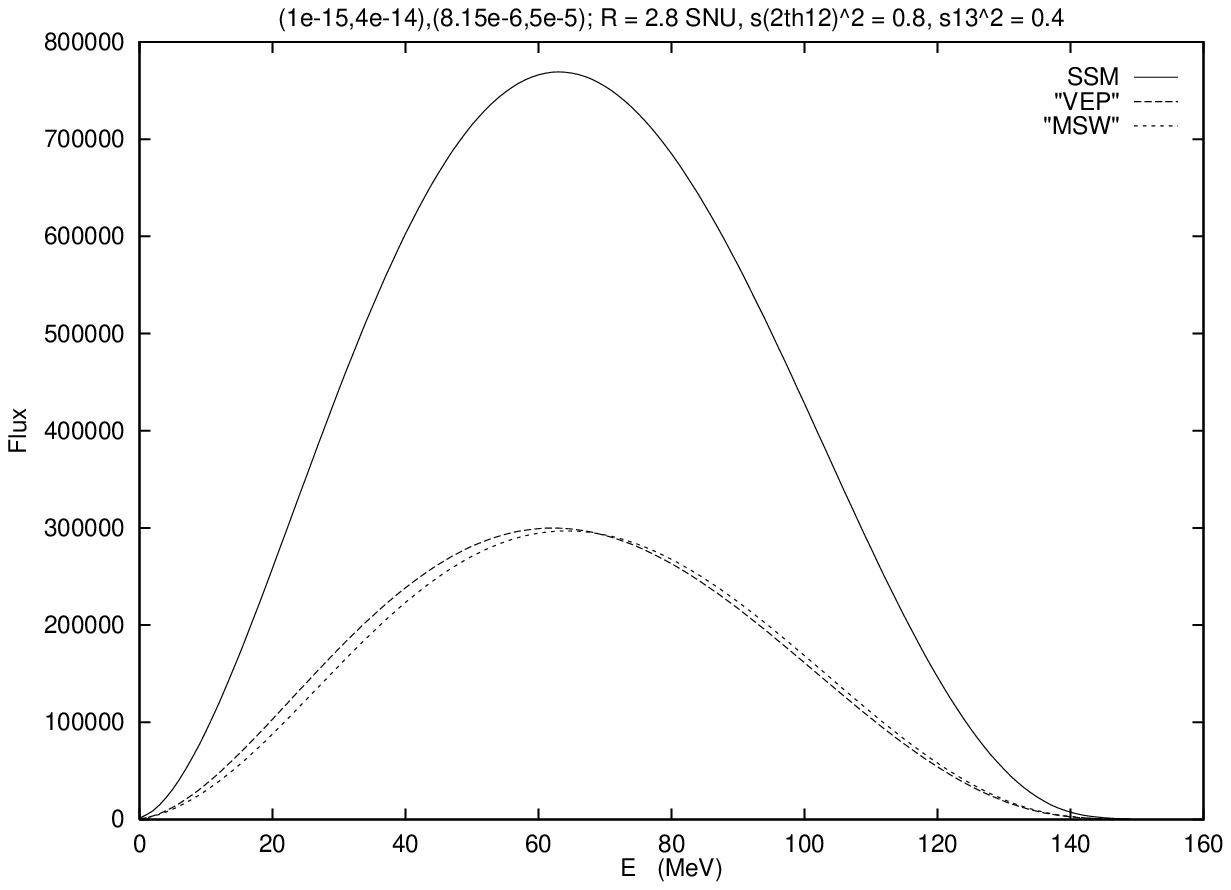}
\caption{}
\label{dll}
\end{figure}

\begin{figure}
\leavevmode
\epsfysize=500pt
\epsfbox[60 20 70 470] {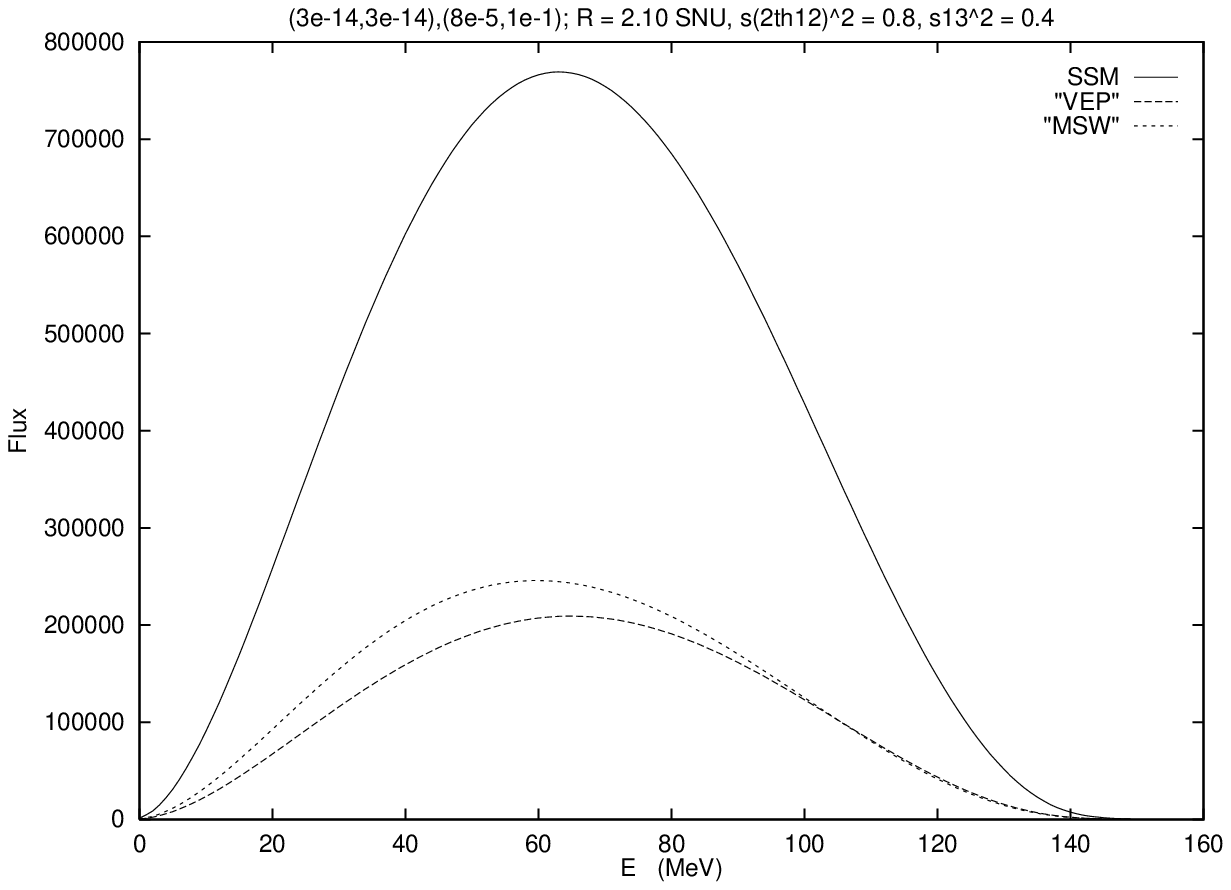}
\caption{}
\label{sll}
\end{figure}

\begin{figure}
\leavevmode
\epsfysize=500pt
\epsfbox[60 20 70 470] {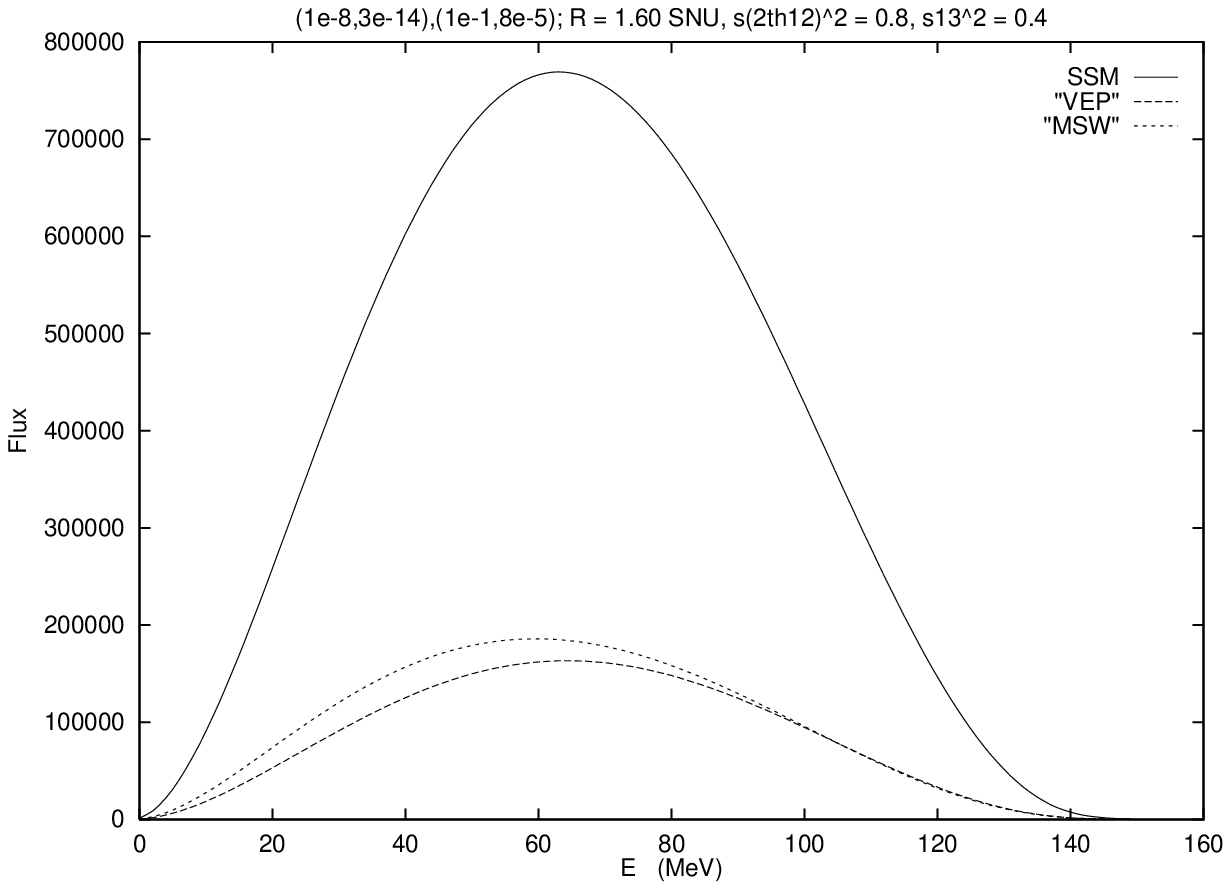}
\caption{}
\label{sllbh}
\end{figure}

\begin{figure}
\leavevmode
\epsfysize=500pt
\epsfbox[60 20 70 470] {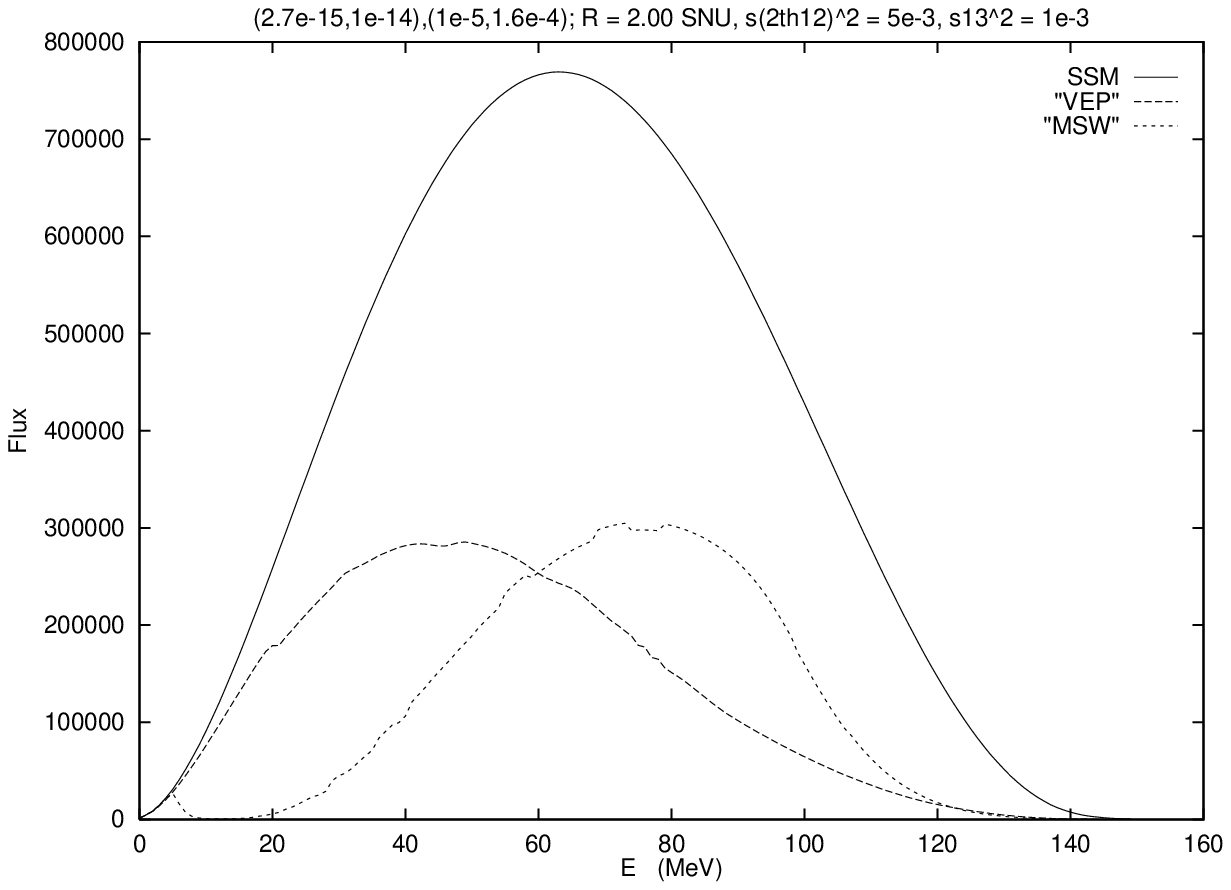}
\caption{}
\label{dss}
\end{figure}

\begin{figure}
\leavevmode
\epsfysize=500pt
\epsfbox[60 20 70 470] {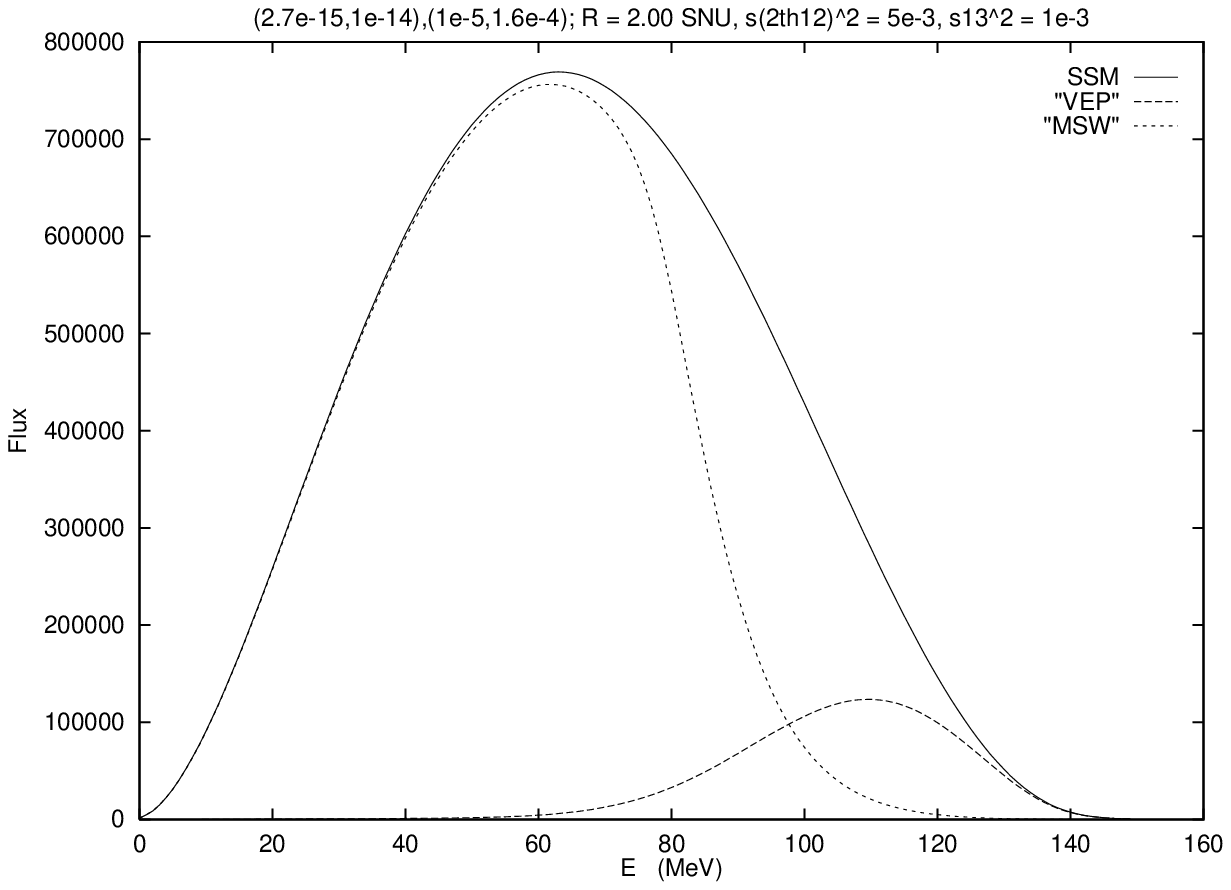}
\caption{}
\label{sssbh}
\end{figure}

	The (chemical detector) counting rate $R^{\alpha}$ for solar
neutrinos from reaction--type
$\alpha$ ({\it e.g.} $\B$, $\Be$ decays, or the pp chain, hereafter
$\B,~\Be,$ and pp neutrinos) is calculated
via
\beq
R^{\alpha} = \int_{0}^{R_{\odot}} \! dr \: r^2 \xi^{\alpha}(r) \int_{E_{min}}^
{E_{max}}
\! dE \: \phi^{\alpha}(E) \sigma(E)\Psurv(r,E)~,
\label{count1}
\eeq
where $\sigma(E)$ is the detector--material absorption cross--section
for neutrinos, $\phi^{\alpha}(E)$ the unreduced neutrino flux \cite{bah3}, and
$\xi^{\alpha}(r)$ the fractional neutrino production rate at radius $r$
\cite{bah2}.
Note that the  size of the flux curve does not reflect the resulting
counting rate, so much as does its {\em shape}. The counting
rate in eq.(\ref{count1}) is essentially calculated as
\beq
R^{\alpha} = \sum_i \phi_i^{\alpha} \sigma_i~,
\label{count2}
\eeq
and so the same $R$ can be obtained for small fluxes as well as for large ones,
since the cross--section $\sigma_i \equiv \sigma(E)$ increases
with increasing energy. It is the location of the unsuppressed
curve on the energy--axis which determines this.
We have performed a numerical integration of Eq.~(\ref{count1})
using the SSM data mentioned in the previous paragraph.
Figures~\ref{dll}--\ref{sssbh}
show both MSW-- and VEP--reduced $\B$ fluxes for various allowed counting
rates, as
compared with the SSM (unreduced) flux of $\Phi^{\B}_{SSM} = 5.8 \times 10^6
\:$cm$^{-2}$
s$^{-1}$.  All mass--differences $\dm_{i1}$
are expressed in units of ~eV$^2$.

The effects of the MSW/VEP mechanisms on solar neutrino depletion
hinge on the existence of a matter--induced flavor--conversion
resonance between different neutrinos ({\it e.g.} $\nue \ra \nm$).
There are three possible resonances for three flavors:
$\nue \ra \nm,~\nue \ra \nt,~ \nm \ra \nt$ (and vice versa).
For solar neutrinos, we are only concerned with flavor conversion
from $\nue \ra \nu_x$, where $\nu_x$ is either of the other two
flavors.\footnote{We do not consider arbitrary sterile neutrinos
in this analysis.}  Resonances can occur if the $\nue$s are created
at an electron density $\ncr > \res_{1i}$, with $i=2,3$ for the
other two flavors, where
\beq
\res_{1i} = \frac{1}{G_F} {\rttwo |\phi(r)| \df E \cos{2\th_{1i}}}
\label{resdens}
\eeq
For MSW, simply make the substition in eq.~(\ref{subs}), and replace
mass--eigenstate vacuum mixing angles with gravitational ones.

Conversely,
if $\ncr < \res_{1i}$ , the $\nue$ will never undergo resonance, and
will propagate as if in vacuum.  We consider here only the case
$\res_{1i} > \nmax\,$, {\it i.e.} the $\nue$ resonance density exceeds
the maximal solar electron density.  We refer to this behavior according
to the following: single resonance ($\ncr > \res_{1i}$ for $i=2$ {\it or}
3 only), and double resonance ($\ncr > \res_{1i}$ for both $i=2,3$).

We note a marked difference
between small and large angle solutions.  Essentially, the large--angle
solutions show very little variation in the fluxes.  This is consistent
with the form of (\ref{lrg}), which show the $\nue$--attenuation
to be energy independent.  The least variation is visible in the
double--resonance
case (fig.~\ref{dll}), while we see more of a discrepancy in the models
for a single--resonance (for both natural-- and broken--hierarchies,
figs~\ref{sll},~\ref{sllbh}).  This difference is mostly in the low--energy
neutrinos, though, and so would be difficult to detect.  The curves
for large $\temu$, small $\tetau$ show similar energy independence,
and thus are not presented here.

	In contrast, the small--angle solution shows very different behaviour.
Figure~\ref{dss} depicts the double--resonance reduced $\B$ neutrino
flux for the two models, with quite surprising dissimilarities.
The two models show {\em opposite}
energy--dependent reduction: MSW suppresses low energy neutrinos, while
VEP suppresses high energy ones\footnote{The rough nature of the curves
is attributed to numerical variations, and most likely not a physical
behavior.}.  Furthermore, the broken hierarchy case
for the small angle region (fig.~\ref{sssbh}) shows other intriguing behavior.
As mentioned earlier, the counting rate is determined by the product of
the cross--section $\sigma(E)$ and the flux $\: \phi(E)$.  Here we see
an example of where both large and small fluxes can  represent the same
counting rate.  Whereas the natural hierarchy attenuated low energy
neutrinos, here we see that the situation is reversed: the large curve
shows the low--energy neutrinos largely uneffected by the MSW mechanism,
while VEP leaves high--energy neutrinos alone.

	This radically different spectral distortion between the two
models can in part be explained by the fact that (especially for small
mixing angles) the adiabatic and non--adiabatic transitions are
reversed.  It is noted in \cite{bah1} that the adiabaticity condition
is violated for low energy neutrinos in VEP, while it is violated for
high energy neutrinos in MSW.  This is clearly reflected in the
small $\tetau$ solutions of figs.~\ref{dss},~\ref{sssbh}.

In summary, the inclusion of a third flavor in each mechanism yields a widely
varying range of possible flux curves depending upon the values of
the mixing angle parameters. The small angle
solution offers variations in the structure of $\Psurv$,
while still approximating the two--flavor limit ($\tetau \ra 0$).
In the case of the small 13--mixing region for double flavor--resonances,
the flux suppression is opposite
for the two mechanisms.  For an unbroken hierarchy, the VEP mechanism
attenuates low energy neutrinos whereas MSW attenuates the higher energy range;
in the case of a broken
hierarchy the roles played by each mechanism are interchanged.
The large $\tetau$ solutions are energy independent, thus making
it difficult to distinguish between MSW and VEP in the high--energy
portion of the spectrum, as most spectral distortions occur here toward
the lower end. Additional information from atmospheric observations
\cite{atm} and laboratory experiments \cite{LSND}
will then be essential in determining the underlying
oscillation mechanism.

\vskip .25 cm

\noindent
{\bf Acknowledgements}

This work was supported in part by the Natural Sciences and Engineering
Research Council of Canada.

\pagebreak
{\large\bf Figure Captions}
All captions list parameters used to obtain the reduced counting
rate $R$.  Neutrino masses are expressed in units of eV$^2$.\\

{\bf Fig. 1}: $\df_{21}=10^{-15}~,~\df_{31}=4 \times 10^{-14}~;
\dm_{21}=6.15 \times 10^{-6}~, \dm_{31} = 5 \times 10^{-5}~;$

{\bf Fig. 2}: $\df_{21}=10^{-15}~,~\df_{31}=4 \times 10^{-14}~;
\dm_{21}=6.15 \times 10^{-6}~, \dm_{31} = 5 \times 10^{-5}~;$

{\bf Fig. 3}: $\df_{21}=10^{-8}~,~\df_{31}=3 \times 10^{-14}~;
\dm_{21}= 10^{-1}~, \dm_{31} = 6 \times 10^{-5}~;$

{\bf Fig. 4}: $\df_{21}=2.7 \times 10^{-15}~,~\df_{31}= 10^{-14}~;
\dm_{21}= 10^{-5}~, \dm_{31} = 1.6 \times 10^{-4}~;$

{\bf Fig. 5}: $\df_{21}=10^{-8}~,~\df_{31}=1.31\times 10^{-14}~;
\dm_{21}= 10^{-1}~, \dm_{31} = 1.31 \times 10^{-4}~;$

\pagebreak

\begin{thebibliography}{99}

\bibitem{clx} R.\ Davis, D.\ S.\ Harmer, and K.\ C.\ Hoffman, Phys.\ Rev.\
Lett.\ {\bf 20}, 1205 (1968).

\bibitem{kii} Kamiokande II Collaboration, K.\ Hirata {\it et al.}, Phys.\
Rev.\ Lett. {\bf 65}, 1297 (1990).

\bibitem{galx} GALLEX Collaboration, P.\ Anselmann {\it et al.}, Phys.\
Lett.\ {\bf B327}, 377 (1994).

\bibitem{sage} SAGE Collaboration, J.\ N.\ Abdurashidov {\it et al.},
Phys.\ Lett.\ {\bf B328}, 234 (1994).

\bibitem{bah2} J.\ N.\ Bahcall and M.\ H.\ Pinsonneault, Rev.\ Mod.\ Phys.\
{\bf 64}, 885 (1992).

\bibitem{msw1} S.\ P.\ Mikheyev and A.\ Yu.\ Smirnov, Yad.\ Fiz.\
{\bf 42}, 1441 (1985).

\bibitem{msw2} S.\ P.\ Mikheyev and A.\ Yu.\ Smirnov, Phys.\ Lett.\
{\bf B21}, 560 (1988).

\bibitem{wolf} L.\ Wolfenstein, Phys.\ Rev.\ {\bf D17}, 2369 (1978).

\bibitem{gasp} M.\ Gasperini, Phys.\ Rev.\ {\bf D38}, 2635 (1988).

\bibitem{bah1} J.\ N.\ Bahcall, P.\ I.\ Krastev, and C.\ N.\ Leung,
IASSNS-AST 94/54, UDHEP-10-94 (October 1994); for earlier work see
M.\ N.\ Butler {\it et al.}, Phys.\ Rev.\ {\bf D47},
2615 (1993); A.\ Halprin and C.\ N.\ Leung, Phys.\ Rev.\ Lett. {\bf 67},
1833 (1991); J.\ Pantaleone, A.\ Halprin, and C.\ N.\ Leung, Phys.\ Rev.\
{\bf D47}, R4199 (1993); K. Iida, H. Minakata and O. Yasuda,
Mod. Phys. Lett. {\bf A8} (1993) 1037.


\bibitem{min} H.\ Minakata and H.\ Nunokawa, KEK-TH-396/TMUP-HEL-9402,
(hep-ph 9405239) (April 1994).

\bibitem{bald} A.\ Baldini and G.\ F.\ Giudice, Phys.\ Lett.\ {\bf B186},
211 (1987).

\bibitem{jrm1} J.\ Mureika, {\it Gravitationally--Induced Three--Flavor
Neutrino Oscillations as a Possible Explanation of the Solar
Neutrino Problem}, M.Sc.\ thesis, University of Waterloo (1995).

\bibitem{zag} H.\ W.\ Zaglauer and K.\ H.\ Schwarzer, Phys.\ Lett.\ {\bf B198}
556 (1987).

\bibitem{bah3} J.\ N.\ Bahcall, {\it Neutrino Astrophysics}, Cambridge
University Press, 1988.

\bibitem{atm}M.C. Goodman, Nucl. Phys. {\bf B38} (Proc. Supp.) (1995)
337; K.S. Hirata {\it et. al.}, Phys. Lett. {\bf B280} (1992)
146; D. Casper {\it et.al.} Phys. Rev. Lett. {\bf 66} (1991) 2561.

\bibitem{LSND}L. Borodovsky {\it et. al.}, Phys. Rev. Lett. {\bf 68}
(1992) 274;
V.V. Ammosov {\it et. al.}, Z. Phys. {\bf C40} (1988) 487.
C. Athanassopoulos {\it et. al.},
Phys. Rev. Lett. {\bf 75} (1995) 2650;  J. Hill,  Phys. Rev. Lett. {\bf 75},
(1995) 2654;
 B.T. Cleveland {\it et. al.}
Nucl. Phys. {\bf B38} (Proc. Supp.) (1995) 47.



\end{thebibliography}
\end{document}